\documentclass{aa}
\usepackage{graphicx}
\usepackage{txfonts}
  \def\beq{\begin{equation}}
\def\eeq{\end{equation}}
\newcommand{\form}[1]{Eq. (\ref{#1})}
\def\etal{et al.}

\begin{document}
   \title{Quantum-Gravity Analysis of Gamma-Ray Bursts using Wavelets}


   \author{J. Ellis\inst{1}
          \and N.E. Mavromatos \inst{2}
          \and  D.V. Nanopoulos \inst{3}
          \and A.S. Sakharov \inst{4}
          }

   \offprints{A.S. Sakharov}

   \institute{Theory Division, CERN, 1211 Geneva 23, Switzerland \\
              \email{John.Ellis@cern.ch}
         \and Department of Physics, Theoretical Physics, King's College London,
          Strand, London WC2R 2LS, England\\ \email{Nikolaos.Mavromatos@cern.ch}
          \and George P. and Cynthia W. Mitchell Institute of Fundamental
 Physics,
Texas A \& M University,
College Station,
TX~77843, USA; \\
Astroparticle Physics Group, Houston Advanced Research Center (HARC),
Mitchell Campus, Woodlands, TX~77381, USA; \\
Academy of Athens,
Division of Natural Sciences, 28~Panepistimiou Avenue, Athens 10679,
Greece\\  \email{Dimitri.Nanopoulos@cern.ch}
\and  Theory Division, CERN, 1211 Geneva 23, Switzerland; \\
    Swiss Institute of Technology, ETH-Z\"urich, 8093 Z\"urich, Switzerland\\
    \email{Alexandre.Sakharov@cern.ch}
}

   \date{Received  29 October 2002 / Accepted 14  February 2003
}

   \abstract{
     In some models of quantum gravity, space-time is thought to have a foamy
structure with non-trivial optical properties.  We probe the possibility
that photons propagating in vacuum may exhibit a non-trivial refractive
index, by analyzing the times of flight of radiation from gamma-ray
bursters (GRBs) with known redshifts. We use a wavelet shrinkage procedure
for noise removal and a wavelet `zoom' technique to define with high
accuracy the timings of sharp transitions in GRB light curves, thereby
optimizing the sensitivity of experimental probes of any energy dependence
of the velocity of light. We apply these wavelet techniques to 64~ms and
TTE data from BATSE, and also to OSSE data. A search for time lags between
sharp transients in GRB light curves in different energy bands yields the
lower limit $M \ge 6.9 \cdot 10^{15}$~GeV on the quantum-gravity scale in
any model with a linear dependence of the velocity of light $\propto E/M$.
We also present a limit on any quadratic dependence.
   \keywords{distance scale --
             gamma ray: bursts --
             methods: statistical
               }
   }

   \maketitle
%

\begin{flushright}
astro-ph/0210124 \\
ACT-02-08, CERN-TH/2002-258, MIFP-02-03 \\
October 2002
\end{flushright}

\section{Introduction}

In standard relativistic quantum field theory, space--time is considered
as a fixed arena in which physical processes take place. The
characteristics of the propagation of light are considered as a classical
input to the theory. In particular, the special and general theories of
relativity postulate a single universal velocity of light $c$. However,
starting in the early 1960s (Wheller \ 1963), efforts to find a synthesis of
general relativity and quantum mechanics, called quantum
gravity, have
suggested a need for greater sophistication in discussing the propagation
of light in vacuum.

A satisfactory theory of quantum gravity is likely to require a drastic
modification of our deterministic representation of space--time, endowing
it with structure on characteristic scales approaching the Planck length
$\ell_{\rm P} \simeq m_{\rm P}^{-1}$. There is at present no complete
 mathematical
model for quantum gravity, and there are many different approaches to the
modelling of space-time foam. Several of these approaches suggest that the
vacuum acquires non-trivial optical properties, because of gravitational
recoil effects induced by the motion of energetic particles. In
particular, it has been suggested that these may induce a non-trivial
refractive index, with photons of different energies travelling at
different velocities. Such an apparent violation of Lorentz invariance can
be explored by studying the propagation of particles through the vacuum,
in particular photons emitted by distant astrophysical
sources (Amelino-Camelia \etal \ 1998). In some quantum-gravity models, light
 propagation
may also depend on the photon polarization (Gambini \& Pullin \ 1999), inducing
birefringence. Stochastic effects are also possible, giving rise to an
energy-dependent diffusive spread in the velocities of different photons
with the same energy (Ford 1995; Ellis \etal \ 2000a).

One may discuss the effects of space--time foam on the phase velocity,
group velocity or wave-front velocity of light. In this paper, we discuss
only the signature of a modification of the group velocity, related to a
non-trivial refractive index $n(E)$: $v(E) = c/n(E)$. This may be derived
theoretically from a (renormalized) effective Maxwell action
$\Gamma_{\rm eff}[{\bf E}, {\bf B}]$, where ${\bf E}$ and ${\bf B}$ are the
electric and magnetic field strengths of the propagating wave, in the
background metric induced by the quantum gravity model under
consideration. Once the effective Maxwell action is known, at least in a
suitable approximation, one can analyze the photon dispersion using the
effective Maxwell equations (Ellis \etal \ 2000b).

One generally considers the propagation of photons with energies $E$ much
smaller than the mass scale $M$ characterizing the quantum gravity model,
which may be of the same order as the Planck mass $M_P$, or perhaps
smaller in models with large extra dimensions. In the approximation $E \ll
M$, the distortion of the standard photon dispersion relation may be
represented as an expansion in $E/M$:

\beq
\label{disprel}
E^2=k^2(1+\xi_1 (k/M)+\xi_2 (k/M)^2+\dots),
\eeq
which implies the following energy dependence of the group velocity
\beq
\label{cquantum}
v\approx c( 1-f(E)).
\eeq
Here the function $f(E)$ indicates the difference of the vacuum refractive
index from unity: $n(E)=1+f(E)$, which is defined by the subleading terms
in the series \form{disprel}.

Different approaches to the modelling of quantum gravity suggest
corrections with different powers of $E/M$. One of the better developed is
a string-inspired model of quantum space--time, in which the corrections
in \form{disprel} start with the third power of $k/M$, as suggested by one
particular treatment of $D$ branes (Ellis \etal \ 1997; Ellis \etal 1998; Ellis
 \etal \ 2000c). In this approach, the
related violation of Lorentz invariance is regarded as spontaneous, and is
due to the impacts of light but energetic closed--string states on massive
$D$(irichlet) particles that describe defects in space--time. In the
modern view of string theory, $D$ particles must be included in the
spectrum, and hence also their quantum fluctuations should be included in
a consistent formulation of the ground--state vacuum. In the model
of Ellis \etal \ 1997; Ellis \etal \ 1998; Ellis \etal \ 2000c, the scattering
 of the closed--string state on the $D$
particle induces recoil of the the latter, which distorts the surrounding
space--time in a stochastic manner, reflecting the foamy structure of
space--time.

In such a picture, the recoil of the massive space--time deffect, during
the scattering with a relativistic low--energy probe such as a photon or
neutrino, distorts the surrounding space--time, inducing an effective net
gravitational field of the form

\beq
\label{gf}
G_{0i}\; \simeq \; \left(\frac{k_i}{M}\right).
\eeq
The dispersion-relation analysis (Ellis \etal \ 2000b) of the Maxwell
equations in the non-trivial background metric perturbed by such a
gravitational field results in a linear dependence of the vacuum
refractive index on the energy:

\beq \label{linearqg}
f(E)\; = \; \left(\frac{E}{M}\right).
\eeq
In some other realisations of quantum gravity, odd powers of $k/M$ in
\form{disprel} may be forbidden (e.g. \ Burgess \etal \ 2002) by selection rules
 such as
rotational invariance in a preferred frame. In this case, the leading
correction to the refractive index takes the form

\beq
\label{quadraticqg}
f(E) \; = \; \left(\frac{E}{M}\right)^2.
\eeq
We assume that the prefactors in both cases are positive, reflecting the
fact that there should be no superluminal propagation (Moore \& Nelson \ 2001).
 This
requirement is not necessarily respected in some models based on the
loop-gravity approach (Gambini \& Pullin \ 1999a; Alfaro \etal \ 2000).

The study of short-duration photon bursts propagating over cosmological
distances is a most promising way to probe this approach to quantum
gravity (Amelino-Camelia \etal \ 1998): for a recent review, see (Sarcar \
 2002). The
modification of the group velocity \form{cquantum} would affect the
simultaneity of the arrival times of photons with different energies.
Thus, given a distant, transient source of photons, one could measure the
differences in the arrival times of sharp transitions in the signals in
different energy bands. Several different types of transient astrophysical
objects can be considered as sources for the photons used to probe
quantum-gravity corrections such as \form{linearqg} and \form{quadraticqg}
to the vacuum refractive index (Amelino-Camelia \etal \ 1998; Ellis \etal \
 2000b; Biller \etal \ 1999; Schafer \ 1999).
These include Gamma-Ray Bursters (GRBs), Active Galactic Nuclei (AGNs) and
pulsars.

A key issue in such probes is to distinguish the effects of the
quantum-gravity medium from any intrinsic delay in the emission of
particles of different energies by the source. Any quantum-gravity effect
should increase with the redshift of the source, whereas source effects
would be independent of the redshift in the absence of any cosmological
evolution effects (Ellis \etal \ 2000b). Therefore, in order to disentangle
source and propagation effects, it is preferable to use transient sources
with a known spread in redshifts $z$. At the moment, one of the most
model-independent ways to probe the time lags that might arise from
quantum gravity is to use GRBs with known redshifts, which range up to $z
\sim 5$.

Increasing numbers of redshifts have been measured in recent years, and
the spectral time lags of GRB light curves have been investigated in a
number of papers (Norris \etal \ 1994; Band \ 1997; Norris \etal \ 2000; Ellis
 \etal \ 2000b; Norris \ 2002). It is important
to detect quantitatively temporal structures which are identical in
different spectral bands, to compare their time positions. Unfortunately,
pulse fitting is problematic (Norris \etal \ 1996; Ellis \etal \ 2000b; Norris
 \etal \ 2000)   in the cases of
many bursts, because of irregular, overlapping structures in the light
curves. As a result these studies often lack the accuracy to characterize
short-time features in the bursts that are evident to the eye. The
 cross-correlation method (Band \ 1997;  Norris \etal \ 2000; Norris \ 2002)
 does not use a rigorous definition of a spike in a pulse; it relies, instead,
 on a calculation of the cross-correlation functions (CCFs) between different
 spectral bands directly in the time domain. However there are some ambiguities
 in the interpretation of CCF peaks, which can lead in some cases to unclear
 conclusions about the spectral evolution. In particular this is the case when a
 GRB light curve contains an emission cluster of closely spaced spikes (e.g.
 spacing of order of the width of a spike); then the width of the CCF's central
 peak, the position of which actually measures the spectral time lag (Band \
 1987), may reflect the duration of the whole cluster and not of the individual
 spikes, whereas only the narrow individual constituents (spikes) of such an
 emission cluster can mark with a good accuracy the  arrival time of radiation,
 so as to apply for a search for quantum gravity effects (Amelino-Camelia \etal
 \ 1998).

In this paper, we seek to overcome the problems mentioned above by using wavelet
transforms to remove noise, to resolve overlapping structures and to
classify quantitatively
the irregularities of GRB light curves with known redshifts. The ability of the
 wavelet technique to characterize burst morphology allows
us to improve significantly the accuracy of the measurements of time lags,
 independently of the degree of spikes separation inside the emission clusters, 
 
increasing the sensitivity to quantum-gravitational corrections. We
analyse the light curves of GRBs with known redshifts triggered by the
Burst And Transient Source Experiment (BATSE) aboard the Gamma Ray
Observatory (GRO) (see the GRO webpage at {\tt
 http://cossc.gsfc.nasa.gov/cgro/index.html}), searching for a redshift
 dependence of spectral time
lags between identical sharp signal transitions detected by wavelet
transforms in different spectral bands. For several GRBs among the
triggers under consideration, one can compare the BATSE light curves with
those measured at higher energies by the Orientated Scintillation
Spectrometer Experiment (OSSE) aboard the GRO.  We also demonstrate that
the wavelet technique can deal with the leading parts of the GRB light
curves recorded by the BATSE time trigger event (TTE), which improves
 the time resolution substantially. Unfortunately, in all the cases except
GRB980329, the TTE data do not cover enough of the light curve to exhibit
coherent structures in different spectral bands, which would increase the
sensitivity to higher quantum-gravity scales.

We find that the combination of all the available data, when analyzed
using wavelet transforms, prefers marginally a linear violation of Lorentz
invariance between $10^{15}$~GeV and $10^{16}$~GeV, although the
effect is not significant. We prefer to interpret the data as giving a
limit on the linear quantum-gravity scale:

\begin{equation}
M \; \ge \; 6.9\cdot 10^{15}~{\rm GeV},
\end{equation}
which we consider to be the most robust and model-independent
currently available.

The layout of this paper is as follows. In Section 2 we discuss the
propagation of light in an expanding Universe, establishing the basic
formulae we use subsequently in our analysis of time lags. The fundamental
definitions and features of wavelet transforms are reviewed in Section 3,
and we describe in Section 4 how wavelet shrinkage can be used to remove
noise from GRB spectra. The `zooming' technique for localizing variation
points in GRB light curves is described in Section 5, and Section 6 uses
this technique to analyze time lags. Our limits on linear and quadratic
quantum-gravity models are obtained in Section 7, and we discuss our
results in Section 8. In addition, Appendix A discusses signal threshold
estimation in the wavelet approach, and Appendix B recalls some aspects of
the Lipschitz characterization of singularities.


\section{Light Propagation in the Expanding Universe}

Light propagation from remote objects is affected by the expansion of the
Universe and depends upon the cosmological model (Ellis \etal \ 2000b).
Present cosmological data motivate the choice of a spatially-flat
Universe: $\Omega_{\rm total} = \Omega_{\rm\Lambda} + \Omega_{\rm M} = 1$ with
cosmological constant $\Omega_{\Lambda} \simeq 0.7$:
see (Bahcall \etal \ 1999)
and references therein. The corresponding differential
relation between time and redshift is
\beq
\label{timez}
dt = - H_0^{-1}\frac{dz}{(1 + z)h(z)},
\eeq
where
\beq
\label{h}
h(z) = \sqrt{\Omega_{\rm\Lambda} + \Omega_{\rm M} (1 + z)^3}.
\eeq
Thus, a particle with velocity $u$ travels an elementary distance
\beq
\label{eldist}
udt=-H_0^{-1}\frac{udz}{(1+z)h(z)},
\eeq
giving the following difference in distances covered by two particles with
velocities differing by $\Delta u$:
\beq
\label{distdiff}
\Delta L = H_0^{-1}\int\limits_0^z\frac{\Delta udz}{(1 + z) h(z)}.
\eeq
We consider two photons traveling with velocities very close to $c$,
whose present day energies are $E_1$ and $E_2$. At earlier epochs, their
energies would have been blueshifted by a factor $1+z$. Defining $\Delta
E \equiv E_2 - E_1$, we infer from equation \form{cquantum} that

\beq
\label{Elin}
\Delta u=\frac{\Delta E(1+z)}{M}
\eeq
in the case \form{linearqg} of a linear $E$-dependence of the velocity of
light, and
\beq
\label{Equadr}
\Delta u=\frac{\Delta E^2(1+z)^2}{M^2}
\eeq
for the quadratic correction \form{quadraticqg}. Inserting the last
two expressions into \form{distdiff}, one finally finds that the induced
differences in the
arrival times of the two photons with energy difference $\Delta E$ are
\beq
\label{timedel1}
\Delta t=H_0^{-1}\frac{\Delta E}{M}\int\limits_0^z\frac{dz}{h(z)},
\eeq
and
\beq
\label{timedel2}
\Delta t=H_0^{-1}\left(\frac{\Delta E}{M}\right)^2\int\limits_0^z
\frac{(1+z)dz}{h(z)}
\eeq
for the linear and quadratic types of correction, respectively.

In the following, we look for such time differences in the arrival times
of photons with energy difference $\Delta E$ propagating in such a flat
expanding Universe with a cosmological constant (Bahcall \etal \ 1999).


\section{What can Wavelet Transforms do?}

Wavelet transforms (WT) (for a review, see Dremin \etal \ 2001) are used to
represent signals which require for their specification not only a set of
typical frequencies (scales), but also knowledge of the coordinate
neighbourhoods where these properties are important. The most important
principles distinguishing a wavelet basis from a windowed Fourier
transform basis are dilatations and translations. Dilatations enable one
to distinguish the local characteristics of the signal at various scales,
and translations enable one to cover the whole region over which the
signal is studied.

The wavelet transform of a function $f$ at the scale $s$ and position $u$
is computed by convoluting $f$ with a wavelet analyzing function:

\beq
\label{wtgeneral}
Wf(u,s)=\int\limits_a^bf(t)\psi_{us}^*(t)dt.
\eeq
The analyzing function $\psi_{us}$ is obtained through dilatation by a
scale factor $s$ and translation by an amount $u$ from a basic (or mother)
wavelet $\psi$:

\beq
\label{mother}
\psi_{us}(t)=\frac{1}{\sqrt{s}}\psi\left(\frac{t-u}{s}\right)
\eeq
It is obvious that $\psi$ must satisfy an admissibility
condition which guarantees the invertibility of the wavelet transform. In
most cases, this condition may be reduced to the requirement that $\psi$
is a function with zero mean (Mallat \ 1998):

\beq
\label{zeroaverage}
\int\limits_{-\infty}^{\infty}\psi (t)dt=0.
\eeq
In addition, $\psi$ is often required to have a certain number of
vanishing moments:

\beq
\label{moments}
\int\limits_{-\infty}^{\infty}t^n\psi (t)dt=0,\qquad n=0,1,\dots ,p.
\eeq
In general, this property improves the efficiency of $\psi$ for detecting
features (singularities) in the signal, since it is blind to polynomials
up to order $N$. One may say that the action of $s$ on the function
$\psi$, which must be oscillating according to \form{zeroaverage}, is a
dilatation if $s>0$ or a contraction if $s<0$. In either case, the shape
of the function is unchanged, it is simply spread out or squeezed.

A transform \form{wtgeneral} over a suitable wavelet basis is usually
called a continuous wavelet transform (CWT). A wavelet
transform whose wavelets $\psi$ are constructed in such a way that the dilated
and translated family
\beq
\label{discretwavelet}
\psi_{j,m}(t)=\frac{1}{\sqrt{2^j}}\psi\left(\frac{t-2^jm}{2^j}\right);\qquad
\eeq
where $j,m$ are integers, is an orthonormal basis for all functions $f$
satisfying the condition
\beq
\label{energy}
\int |f|^2(t)dt<+\infty .
\eeq
This is called a discrete wavelet transform (DWT): for a review,
see (Dremin \etal \ 2001; Mallat \ 1998).

   \begin{figure*}
   \centering
   \includegraphics[width=16cm]{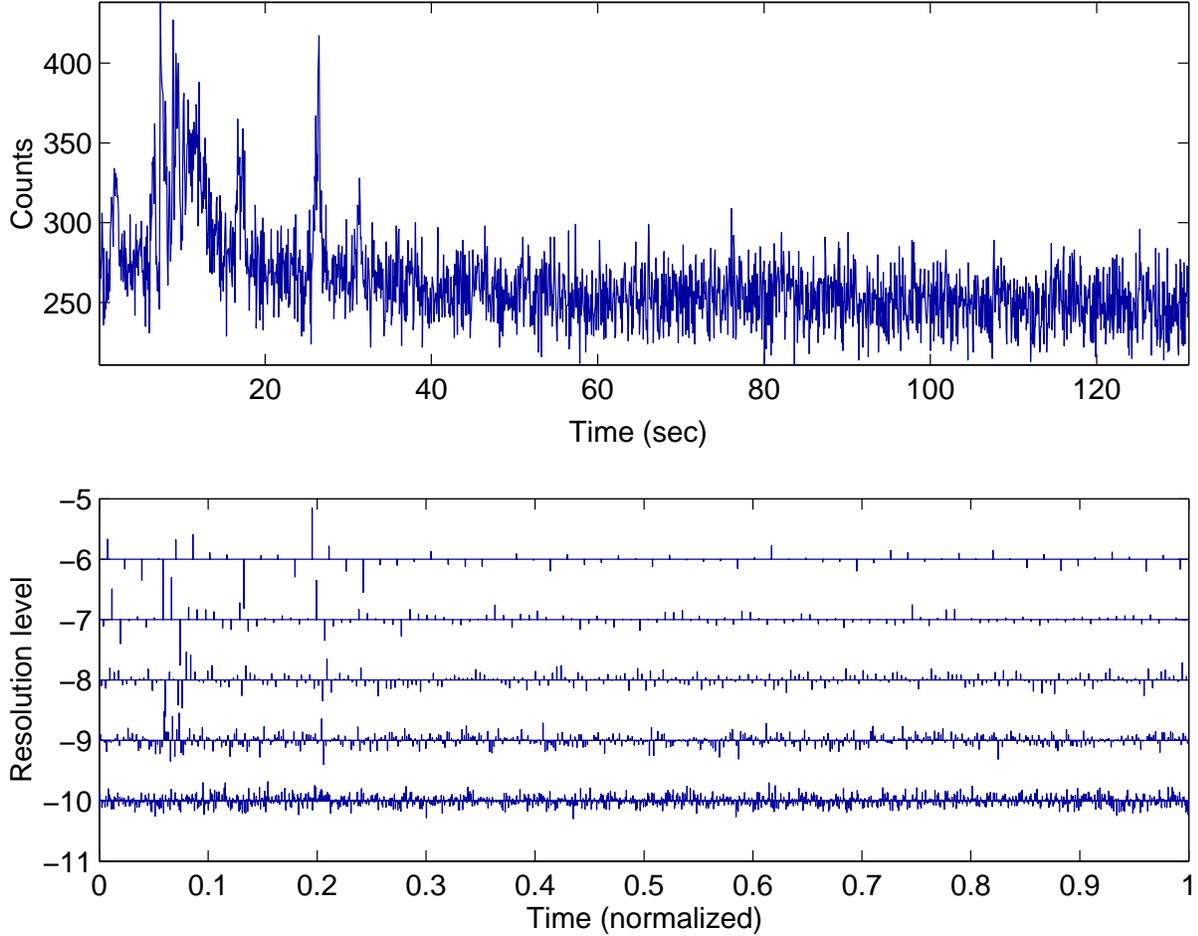}
   \caption{Data from GRB990308 with $z=1.60$ collected by BATSE trigger
7457 in a total number of $2^{11}$ $64$~ms bins in the energy
band 115-320~keV (top panel), and its Symmlet-10 (Mallat \ 1998) discrete
wavelet
transform (DWT) at the level $L=6$. The horizontal axis corresponds to the
same time scale as in the original burst. Each level on the vertical axis
shows the wavelet coefficients at a given resolution level (i.e., scale).
The wavelet coefficients are represented by spikes whose size and direction
(up or down) is determined by the magnitude and sign (+ or -) of the
coefficient.}
   \label{fig:distribution}
    \end{figure*}

The CWT is mostly used for the analysis and detection of signals, whereas
the DWT is more appropriate for data compression and signal
reconstruction. Combining these two types of wavelet transforms provides
an advanced technique for picking up the positions of particular breaks in
the structures of observed GRB light curves in different energy bands,
which we use here to look for non-trivial medium effects on the
propagation of photons due to quantum gravity.

Orthogonal wavelets \form{discretwavelet} dilated by factors $2^j$ are
sensitive to signal variations with resolutions $2^{-j}$. This property
can be used to make a sequence of approximations to a signal with
improving resolutions (e.g. \  Mallat \ 1998). For a function satisfying the
condition \form{energy}, the partial sum of wavelet coefficients
$\sum_{k=-\infty}^{+\infty}d_{j,k}\psi_{j,k}$ can be interpreted as the
difference between two approximations to $f$ with resolutions $2^{-j+1}$
and $2^{-j}$. Adapting the signal resolution allows one to process only
the details relevant to a particular task, namely to estimate intensity
profiles of GRB light curves preserving the positions of sharp signal
transients.

CWTs can detect with very high precision the positions where the intensity
profile of a GRB light curve, as estimated by a DWT, changes its degree of
regularity. Since $\psi$ has zero average, a wavelet coefficient $Wf(u,s)$
measures the variation of $f$ in a neighborhood of $u$ whose size is
proportional to $s$. Sharp signal transitions create large-amplitude
wavelet coefficients. As we see in the following section, the pointwise
regularity of $f$ is related to the asymptotic decay of the wavelet
transform $Wf(u,s)$ when $s$ goes to zero. Singularities are detected by
following across different scales the local maxima of the wavelet
transform. We use this `zooming' capability to define the positions of
mathematically similar transients (irregularities) in GRB light curves
observed in different energy bands. These therefore provide the best
information about the arrival times of photons associated with universal
intrinsic emission features at the sources.

\section{Extraction of the GRB Intensity Profiles by Wavelet Shrinkage}
   The observed GRB light curves typically feature a relatively homogeneous,
nonzero background intensity, above which some inhomogeneous structure is
apparent (Kolaczyk \ 1997). In the following, we demonstrate that when such a
temporally inhomogeneous signal as the light curve of a GRB contains both
structure and noise, the ability of the DWT to compress the information in
this signal leads efficiently to a simple but effective noise removal
procedure. This wavelet shrinkage technique (Donoho \ 1993; Donoho \etal \
 1995), based on the
thresholding of the DWT, allows one to separate the structure of the
signal from the noise, whilst retaining information about the position of
irregularities of the signal, as provided by the support of the mother
wavelets.

 \begin{figure*}
   \centering
   \includegraphics[width=16cm]{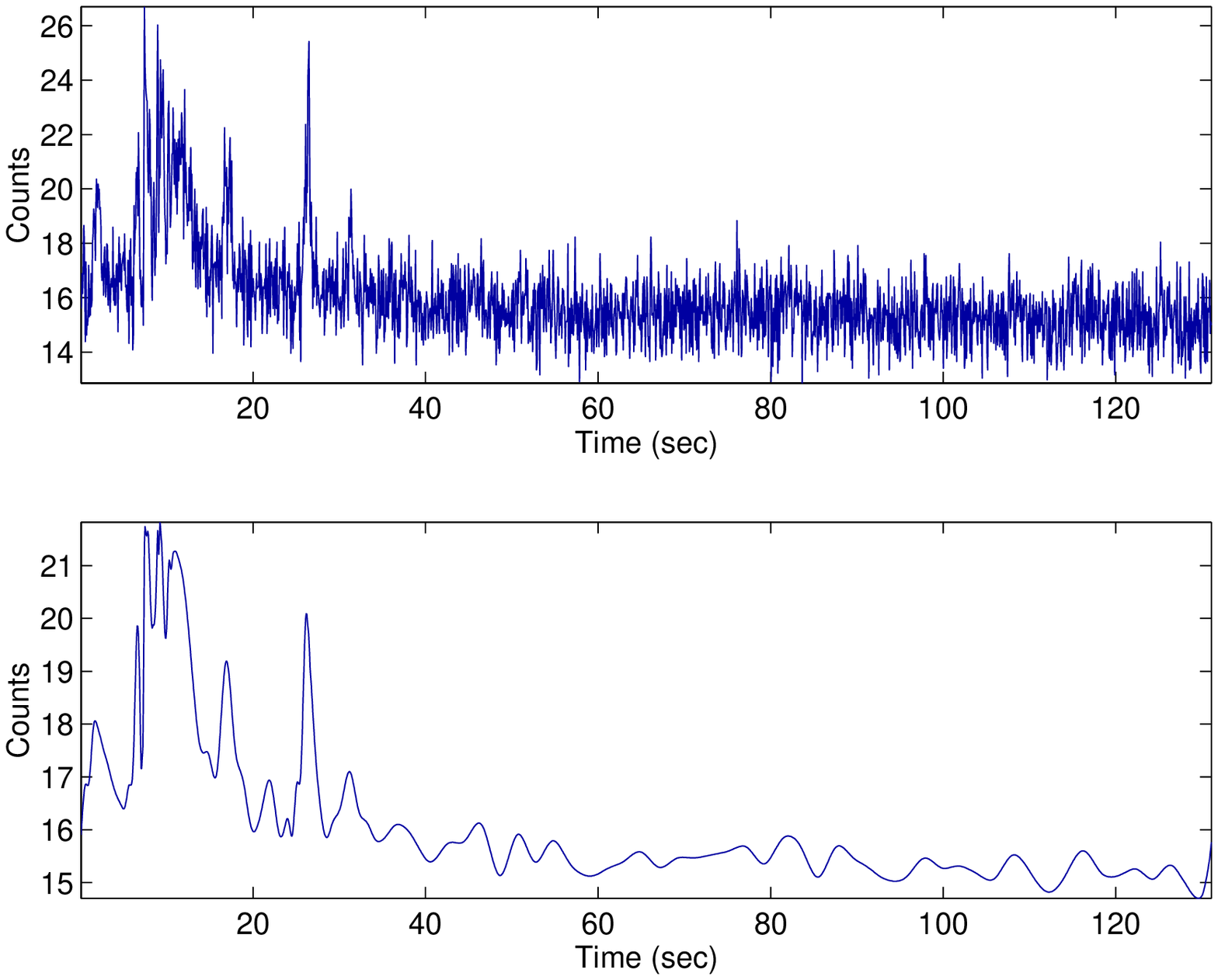}
   \caption{The GRB990308 light curve after preprocessing which sets the
median absolute deviation of wavelet coefficients at the fine scale equal to
unity, as described in more detail in Appendix~\ref{apa}. The intensity
profile estimated by the wavelet shrinkage procedure at
the level $L=6$ is presented in the bottom panel.}
   \label{smoothed}
    \end{figure*}

 In practice, DWTs break a function down into a coarse approximation at a
given scale, that can be extended to successive levels of residual detail
on finer and finer scales. The full decomposition may be expressed in
terms of the scale function $\phi_{j,n}$ and the discrete wavelet
$\psi_{j,n}$ discussed already. The scale function looks much like a
kernel function, and a finite linear combination of dyadic shifts of this
function provides a generic coarse approximation. Further linear
combinations of dyadic shifts of the wavelet function supply the residual
detail. By considering sequences of DWTs with increasing numbers of dyadic
dilatations, the detail at each of the corresponding successive scales is
recovered.

We represent the light curve of a given GRB by a binned discrete signal
$\{X_0,X_1,\dots ,X_{n-1}\}$ of dyadic length $n=2^J$, where $J>0$. The
DWT of such a signal results in a vector of length $n$ of wavelet
coefficients.  The signal is said to have been sampled at level $J$. At
some coarser resolution level (scale) $L< J$, the wavelet coefficient
vector contains $2^L$ scale coefficients $c_{L,0},\dots ,c_{L,2^{L-1}}$
and $2^j$ detail coefficients $d_{j,0},\dots ,d_{j,2^{j-1}}$ at each of
the levels $j=L,\dots ,J-1$. Fig.~\ref{fig:distribution} displays a
typical example with $J=11$ and $L=6$.

Observations of a given GRB light curve can be represented by the sum
\beq
\label{contamination}
X[n]=f[n]+W[n],
\eeq
where the intensity profile $f[n]$ is contaminated
by the addition of noise, which is modelled as a realization $W[n]$
of a random process whose probability distribution is known.
The intensity profile $f$ is estimated by transforming the noisy data
$X[n]$ with the `decision operator' $D$:
\beq
\label{estimator}
\tilde F=DX.
\eeq
The `risk' of the estimator $\tilde F$ of $f$ is the average loss,
calculated with respect to the probability distribution of the noise. The
numerical value of the risk is often specified by the signal-to-noise
ratio (SNR), which is measured in decibels.

To reduce the noise level of $W$, while preserving the degree of
regularity of the intensity profile $f$, we use a soft thresholding
procedure. This procedure sets to zero all coefficients smaller in
magnitude than some threshold $T$, and shrinks coefficients larger than
$T$ towards zero by amounts $T$, as described in more detail in
Appendix~\ref{apa}. This performs an adaptive smoothing that depends on
the regularity of the signal $f$. In a wavelet basis~\footnote{In general,
the thresholding procedure can be applied to any basis for the signal
decomposition (e.g. \ Mallat \ 1998).} where large-amplitude coefficients correspond
to transient signal variations, this means that the estimator discussed in
Appendix~\ref{apa} keeps only transients coming from the original
signal, without adding others due to the noise. After the preprocessing,
which sets the median value of wavelet coefficients of the signal at the
finest scale to unity, the threshold is estimated to be $T=\sqrt{2\log
n}$. The example of an intensity profile estimated by this wavelet
shrinkage procedure, as described in Appendix~\ref{apa}, is shown in
Fig.~\ref{smoothed}.

In general, the wavelet shrinkage procedure \form{softthreshold} described
above guarantees with high probability that $|d_{j,m}^{\tilde F}|\le
|d_{j,m}^ f|$ (e.g. \ Mallat \ 1998), implying that the estimator $\tilde F$ is at
least as regular as the `original' intensity profile $f$, because its
wavelet coefficients have smaller amplitudes. Thus we use this property of
the DWT of separating very effectively the structures in the GRB intensity
profiles from noise, in the form of two subsets of wavelet coefficients,
large and small ones. The thresholding procedure deletes wavelet
coefficients below the threshold value, and diminishes the others by the
threshold value. This tends to preserve both broad and narrow features,
while significantly reducing noise fluctuations, after the reconstruction
of the intensity profile by the inverse DWT.

\section{Detection of Variation Points in GRB Light Curves}

 \begin{figure*}
   \centering
   \includegraphics[width=16cm]{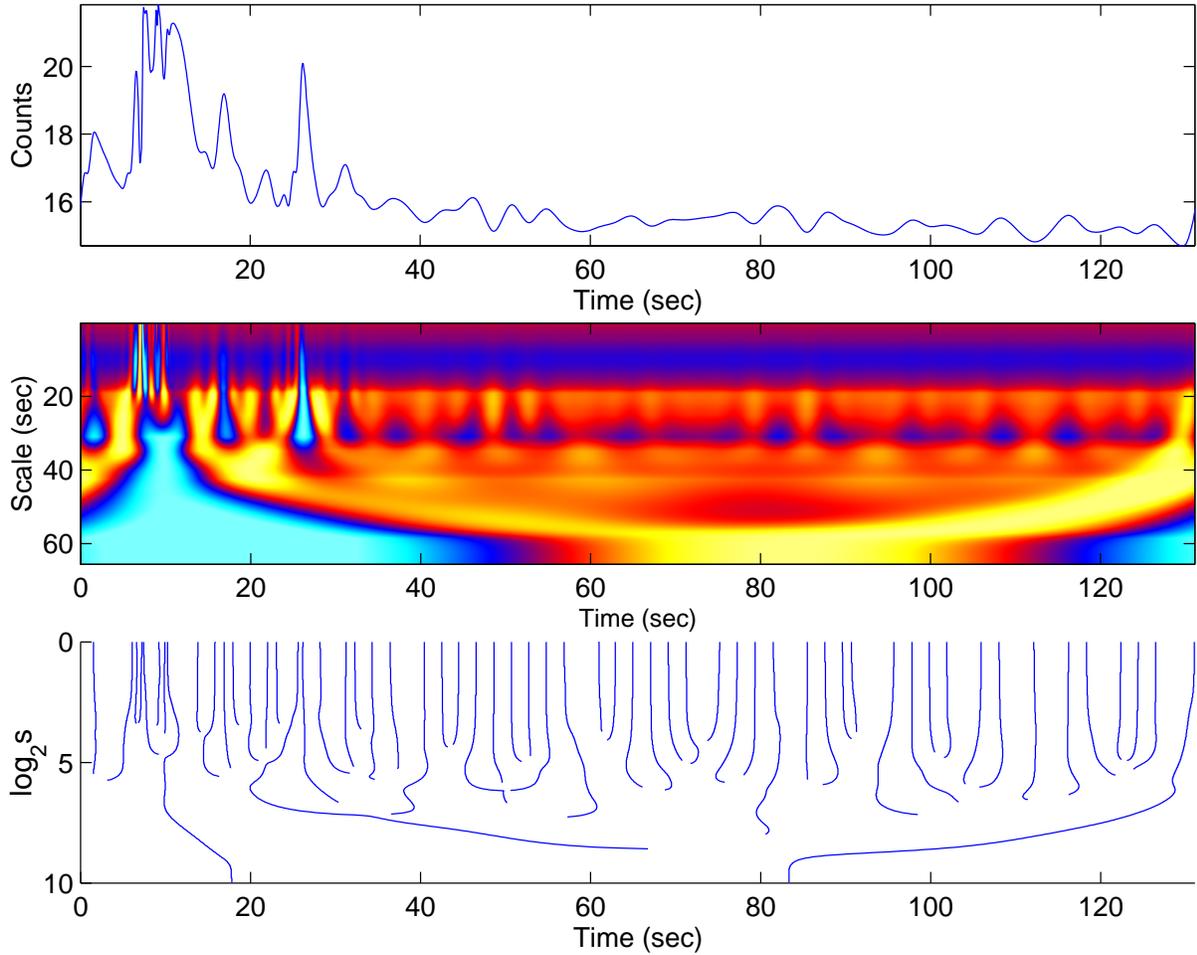}
   \caption{The CWT image (middle panel) of the GRB990308 intensity profile (top
panel). The horizontal and vertical axes give, respectively, the position
$u$ and $\log_2s$ (where $s$ is the time scale in seconds),
The shadings (colours
from yellow to blue) correspond to negative, zero and positive wavelet
transforms respectively. Singularities create large amplitude coefficients in
their cone of influence. The modulus maxima (bottom panel) of $Wf(u,s)$ obtained
from the matrix of CWT (middle panel) pointing towards the time positions of
singularities at the fine scale.}
   \label{cwtimage}
    \end{figure*}

\begin{figure*}
   \centering
   \includegraphics[width=16cm]{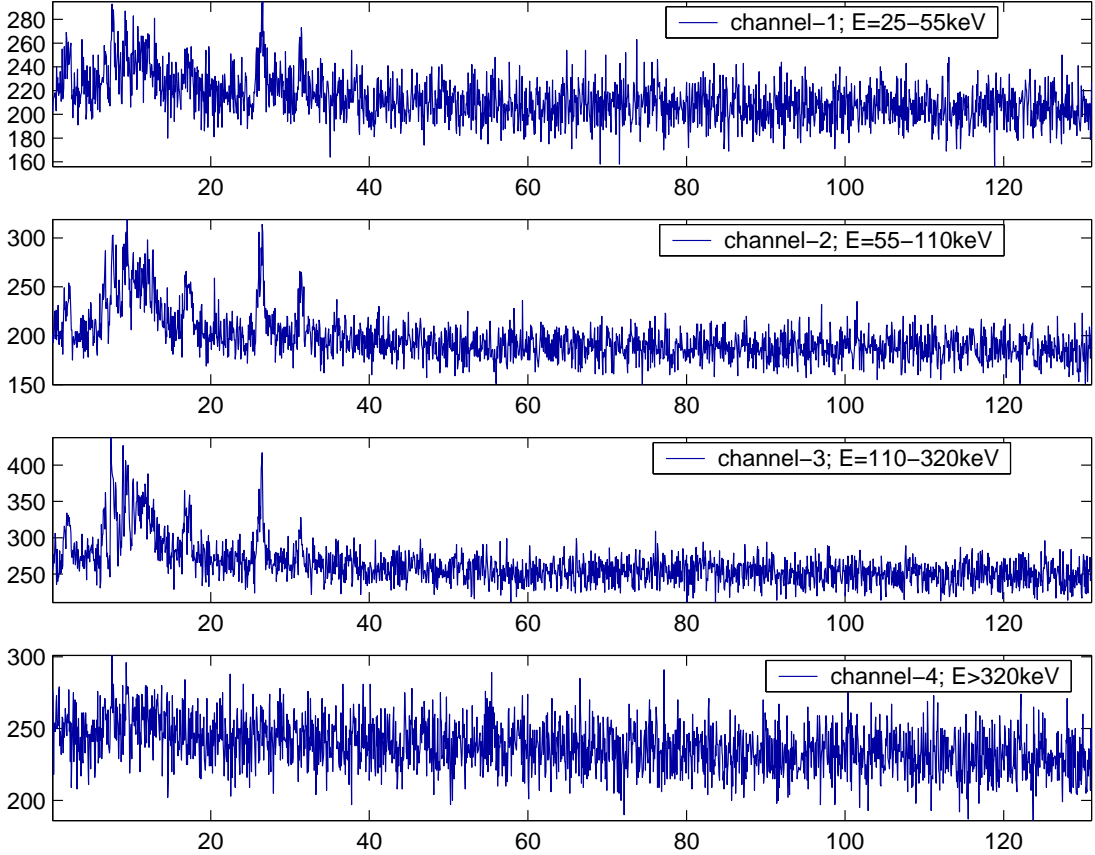}
   \caption{The light curve for GRB990308 obtained by BATSE with trigger 7457, binned
with 64~ms resolution in four spectral bands.}
   \label{batse4}
    \end{figure*}

\begin{figure*}
   \centering
   \includegraphics[width=16cm]{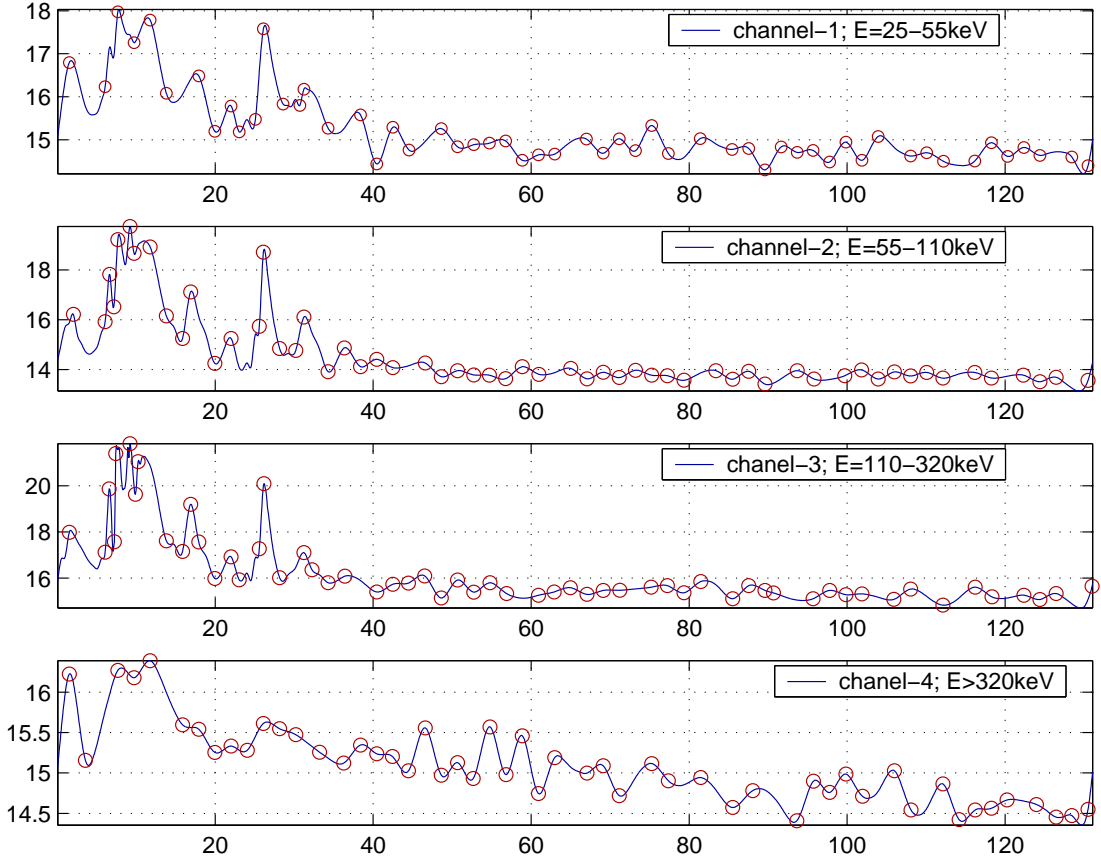}
   \caption{The estimated intensity profiles of GRB990308 - see Fig.~\ref{batse4} -
obtained in four spectral bands by using a Symmlet-$10$ (Mallat \ 1998) basis
at level $L=6$. The signal-to-noise ratios are at the levels $SNR1=23.07$,
$SNR2=22.71$, $SNR3=23.58$ and $SNR4=23.87$ for each band, respectively.
All variation points founded by CWT zoom are marked by circles. The
behaviours of the Lipschitz exponents $\alpha$ are estimated. Seven
pairs of genuine variation points in the first and third spectral bands
have been detected.}
   \label{alldetect}
    \end{figure*}

\begin{figure*}
   \centering
   \includegraphics[width=16cm]{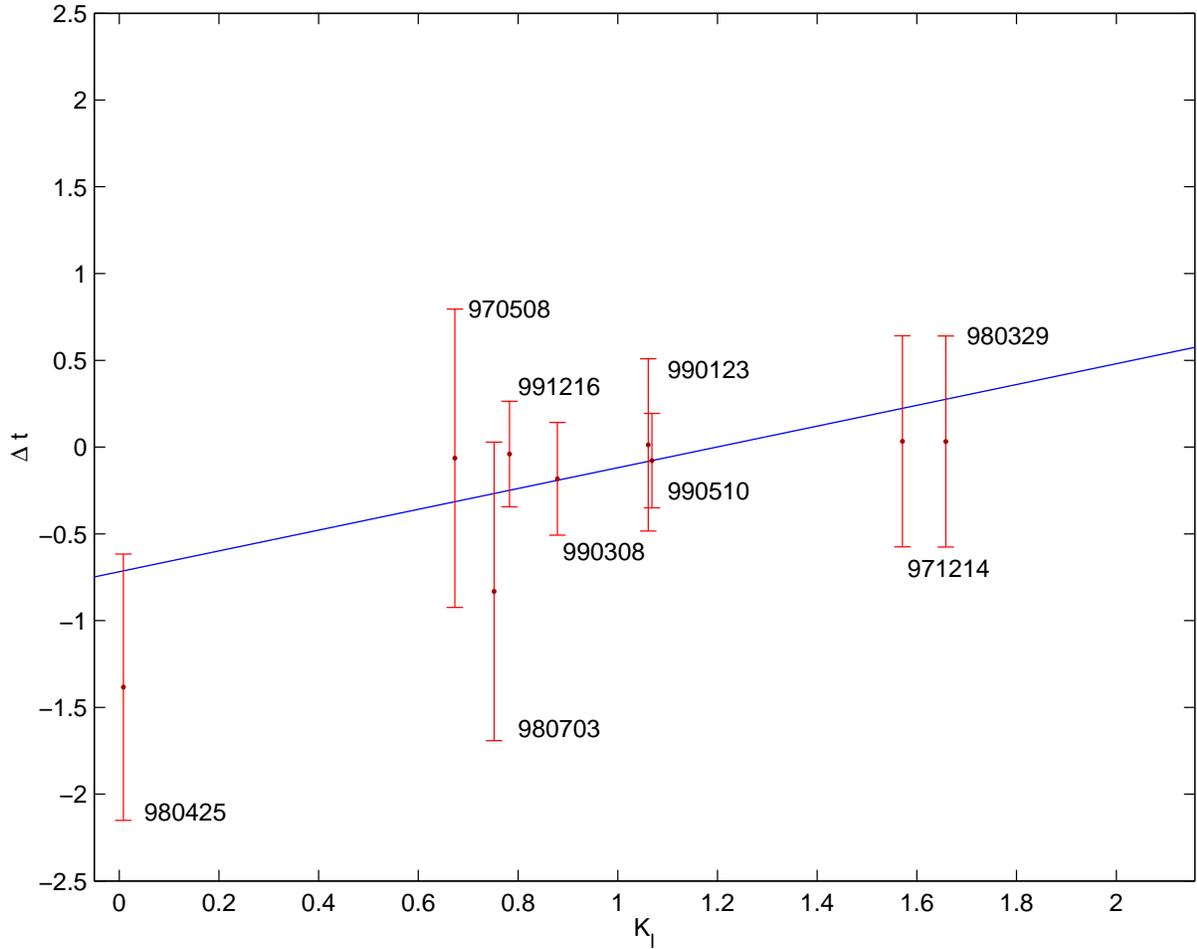}
   \caption{Spectral time lags between the arrival times of pairs of genuine variation
points detected in the third and first BATSE spectral bands. The
analysis has been done for 9 GRB light curves collected with time
resolution 64~ms. The solid line shows the best linear fit versus $K_{\rm l}$.}
   \label{regr64}
    \end{figure*}

\begin{figure*}
   \centering
   \includegraphics[width=16cm]{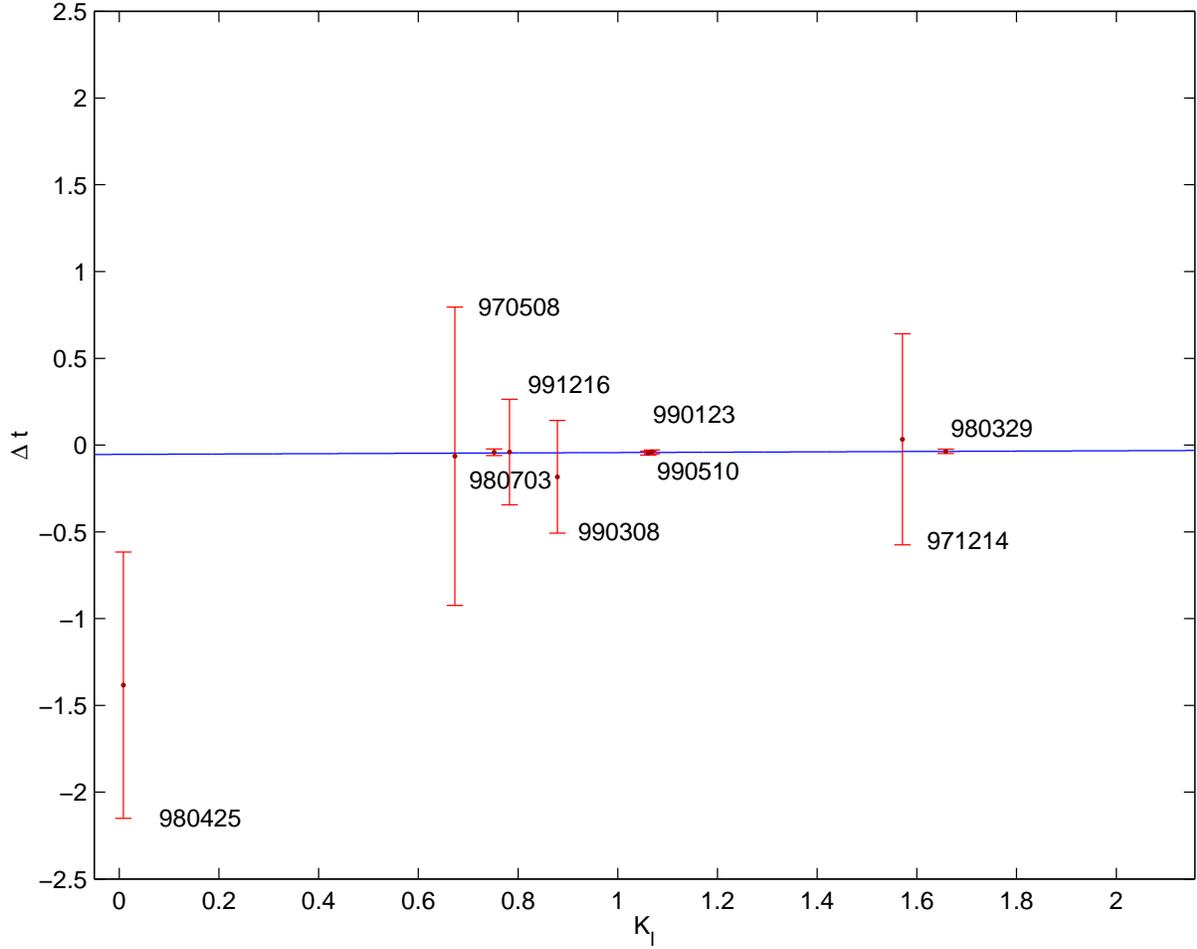}
   \caption{The combination of $64$~ms BATSE time-lag measurements shown in
Fig.~\ref{regr64} with the measurement obtained from the BATSE-OSSE
comparison and the TTE portion of the GRB980329 light curve, with
resolution 2.7~ms.}
   \label{regrtte}
    \end{figure*}

\begin{figure*}
   \centering
   \includegraphics[width=16cm]{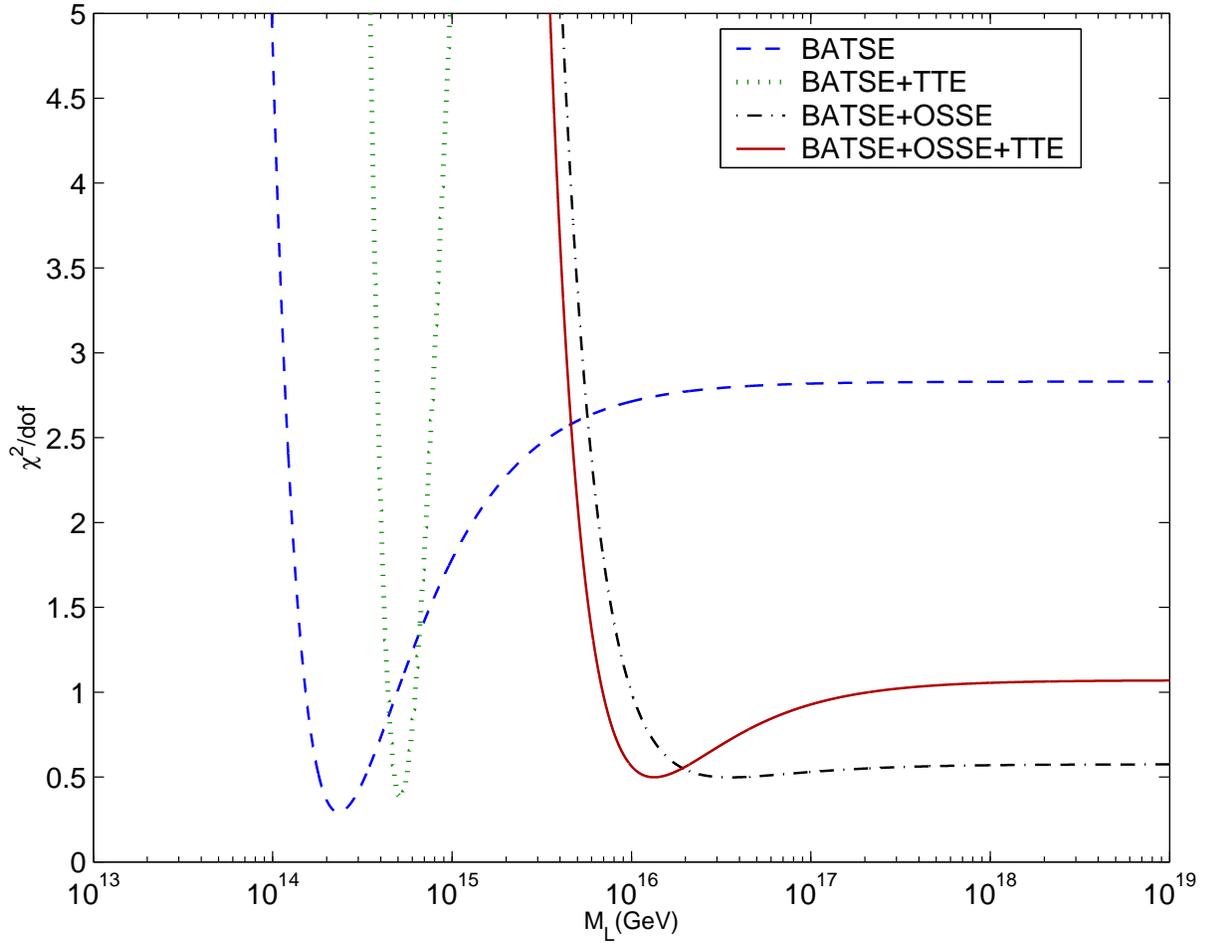}
   \caption{The $\chi^2$ function defined in \form{chi2} as a function of the quantum 
gravity scale, for
the scenario with a linear energy dependence of the vacuum refractive 
index, as obtained using
different combinations of data sets. The solid line, which is used to
establish the lower limit \form{linlimit}, corresponds to the combination 
of $64$~ms
BATSE data with both TTE and OSSE-BATSE data.}
   \label{likelihood}
    \end{figure*}

\begin{figure*}
   \centering
   \includegraphics[width=16cm]{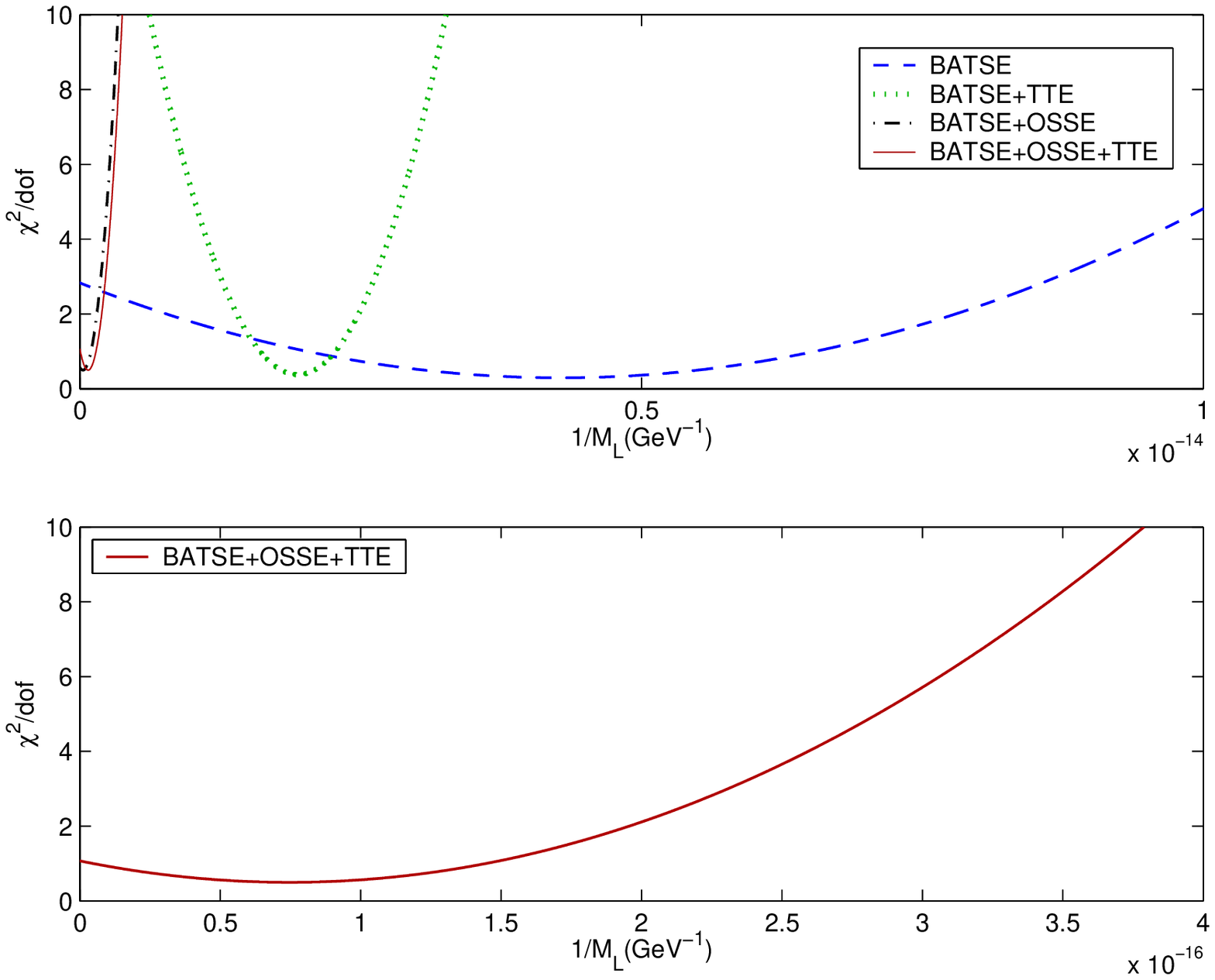}
   \caption{The same $\chi^2$ function as in Fig. \ref{likelihood}, which is
defined in \form{chi2}, now plotted as a function of the inverse
quantum-gravity scale. The solid line corresponds the final lower limit
\form{linlimit}.}
   \label{likelihoodinv}
    \end{figure*}

A wavelet transform can focus on localized signal structure via a
`zooming' procedure that reduces progressively the scale parameter. To
identify the variability points of two different light curves and
characterize their structures, it is necessary to quantify precisely the
local regularity of the function which represents the intensity profile of
the original signals. The appropriate tools are the Lipschitz
exponents~\footnote{Lipschitz exponents are also called H\"older exponents
in the mathematical literature (Dremin \etal \ 2001; Mallat \ 1998).}. These are defined in
Appendix~\ref{apb}, and can provide uniform regularity measurements of a
function $f$ not only over time intervals, but also at any point $\nu$.

We use the Lipschitz exponent $\alpha$ to characterize variation points of
the reconstructed intensity profiles of GRBs accumulated in different
energy bands. The comparison of the positions of variation points with the
same values of $\alpha$ gives the information about the arrival times of
photon probes with different energies, enabling one to probe for
quantum-gravity time-delay phenomena.

The decay of the CWT amplitude as a function of scale is related to the
uniform and pointwise Lipschitz regularity of the signal. Thus, measuring
this asymptotic decay is equivalent to `zooming' into signal structures
with a scale that goes to zero. Namely, when the scale $s$ decreases, the
CWTs $Wf(u,s)$ \form{wtgeneral} measures fine-scale variations in the
neighborhood of $u$. One can prove (e.g. Mallat \ 1998) that $|Wf(u,s)|$ decays
like $s^{\alpha +1/2}$ over intervals where $f$ is uniformly Lipschitz
$\alpha$. Furthermore, the decay of $|Wf(u,s)|$ can be controlled from its
local maxima values.

A `modulus maximum' (e.g. Mallat \ 1998) is any point $(u_0,s_0)$ such
that $|Wf(u,s_0)|$ is locally maximal at $u=u_0$. This local maximum
should be a strict local maximum in either the right or the left
neighborhood of $u_0$. Any connected curve $s(u)$ in the scale-time plane
$(u,s)$ along which all points are modulus maxima, as illustrated in
Fig.~\ref{cwtimage}, is called a `maxima line'. Singularities of a
function $f$ are detected by finding the abscissa where the wavelet
modulus maxima converge on fine scales (e.g. Mallat \ 1998). Only at such points
can $f$ be singular, i.e., with exponent $\alpha\le 1$. This result
guarantees that all singularities are detected by following the wavelet
transform modulus maxima at fine scales. Fig.~\ref{cwtimage} shows an
example where all the significant singularities are located by following
the maxima lines. The positions of these singularities are located by the
modulus maxima lines at the fine scale of decomposition.

To be sensitive to both sharp and smooth singularities, one has to use
wavelets with two vanishing moments, so as to generate the CWT
\form{wtgeneral} of the reconstructed intensity profiles. The most
suitable one is the second derivative of a Gaussian (Mexican hat) mother
wavelet:
\beq
\label{mexhat}
\Psi (t)=\frac{2}{\pi^{1/4}\sqrt{3\sigma}}\left(\frac{t^2}{\sigma^2}-1\right)
\exp\left( -\frac{t^2}{2\sigma^2}\right),
\eeq
because of the property that the modulus maxima of $Wf(u,s)$ with the
wavelet \form{mexhat} belong to connected curves that are not broken as
the scale $s$ decreases (e.g. Mallat \ 1998), which guarantees that all maxima
lines propagate to the finest scales. The dilatation step $s$ is generally
set to $s=2^{1/\Delta}$, where $\Delta$ is the number of intermediate
scales (voices) for each octave. Thus, if the voice lattice is
sufficientely fine, one can build maxima lines with very high precision.
Connecting maxima into lines as in Fig.~\ref{cwtimage} is a procedure for
removing spurious modulus maxima created by numerical errors in regions
where the CWT is close to zero.

Sometimes the CWT may have a sequence of local maxima that converge to a
point $\nu$ on the abscissa, even if $f$ is perfectly regular at $\nu$.
Thus, to detect singularities it is not sufficient merely to follow the
wavelet modulus maxima across the scales. One must also calculate the
Lipschitz regularity from the decay of the modulus maxima amplitude. If,
for some scale $s < s_0$, all the modulus maxima that converge to $\nu$
are included in a cone:

\beq
\label{cone}
|u-\nu |\le Cs,
\eeq
then $f$ has an isolated singularity at $\nu$. Conversely, the absence of
maxima below the cone of influence \form{cone} implies that $f$ is uniform
in the neighborhood of any point $t\neq\nu$ beyond the scale $s_0$.

The Lipschitz regularity at $\nu$ is given by the slope $\log_2|Wf(u,s)|$
as a function of $\log_2s$ along the maxima lines converging to $\nu$,
namely

\beq
\label{wtdecay}
\log_2|Wf(u,s)|\approx\left(\alpha+\frac{1}{2}\right)\log_2s+const.
\eeq
Actually, the Lipschitz property \form{lip} approximates a function with a
polynomial $p_{\nu}$ in the neighborhood of the point $\nu$. The CWT
estimates the Lipschitz exponents of the function by ignoring the
polynomial $p_{\nu}$ itself. Moreover, if the scale $s_0$ is smaller then
the distance between two consecutive singularities, to avoid having other
singularities influence the value of $Wf(u,s)$, and the estimated
Lipschitz exponent $\alpha +1/2\le 1.5$, the function $f$ exhibits a break
at $\nu$, which can be detected by following the modulus maxima chain.

In this paper, to define significant points in the time series of the
signal, we do not apply fit functions that select only prominent peaks, as
was done in (Ellis \etal \ 2000b; Norris \etal \ 1995). We consider a more general class of
relatively sharp signal transitions, marked by Lipschitz irregularities
picked out by the CWT `zoom' technique, which we call {\it genuine
variation points}.
\begin{table*} [t]
\begin{center} \begin{tabular}{|c|c|c|}
\hline
{\bf GRB (BATSE Trigger)}&
{\bf $z$}&
{\bf Time lag, error (s)}\\
\hline
\hline
970508 (6225)&
0.835& {\bf BATSE (64~ms)}\\
&&$\Delta t_1^{\rm BATSE}=-0.064$ $\pm 0.860$\\
\hline
971214 (6533)&
3.418&
{\bf {\rm BATSE} (64~ms)}\\
&&
$\Delta t_1^{\rm BATSE}=0.13$ $\pm 0.860$;
 $\Delta t_2^{\rm BATSE}=-0.064$ $\pm 0.860$\\
&& {\bf cobmbined BATSE (64~ms)}\\
&&$\Delta t_{\rm comb}^{{\rm BATSE}}=0.033$ $\pm 0.608$\\
\hline
980329 (6665)&
$3.9$ &  {\bf BATSE (64ms)}\\
&& $\Delta t_1^{{\rm BATSE}}=0.064$ $\pm 0.860$; $\Delta t_2^{{\rm BATSE}}=0$ $\pm 0.860$\\
&& {\bf combined BATSE (64~ms)}\\
&& $\Delta t_{\rm comb}^{{\rm BATSE}}=0.032$ $\pm 0.608$\\
&& {\bf BATSE-TTE (2.7~ms)}\\
&& $\Delta t_1^{\rm TTE}=-0.34$ $\pm 0.019$\\
&& {\bf OSSE (64~ms) rescaled}\\
&& $\Delta t_1^{{\rm OSSE}}=-0.048$ $\pm 0.019$;
 $\Delta t_2^{{\rm OSSE}}=0$ $\pm 0.038$\\
&& {\bf combined BATSE+OSSE+TTE}\\
&& $\Delta t_{\rm comb}^{{\rm BATSE+rm OSSE+TTE}}=-0.036$ $\pm 0.013$\\
\hline
980425 (6707)&
0.0085& {\bf BATSE (64~ms)}\\
&& $\Delta t_1^{{\rm BATSE}}=-1.792$ $\pm 1.705$; $\Delta t_2^{{\rm BATSE}}=-1.28$ $\pm 0.860$ \\
&& {\bf combined BATSE (64~ms)}\\
&& $\Delta t_{\rm comb}^{{\rm BATSE}}=-1.384$ $\pm 0.768$\\
\hline
980703 (6891)&
0.966& {\bf BATSE (64~ms)}\\
&& $\Delta t_1^{{\rm BATSE}}=-0.832$ $\pm 0.860$\\
&& {\bf OSSE (64~ms) rescaled}\\
&& $\Delta t_1^{{\rm OSSE}}=-0.040$ $\pm 0.019$\\
&& {\bf combined BATSE+OSSE}\\
&& $\Delta t_{\rm comb}^{\rm BATSE+ OSSE}=-0.041$ $\pm 0.019$\\
\hline
990123 (7343)&
1.600& {\bf BATSE (64~ms)}\\
&& $\Delta t_1^{{\rm BATSE}}=0.230$ $\pm 0.860$;
 $\Delta t_2^{{\rm BATSE}}=-0.064$ $\pm 0.860$;
 $\Delta t_3^{{\rm BATSE}}=-0.128$ $\pm 0.860$\\
&& {\bf combined BATSE (64~ms)}\\
&& $\Delta t_{\rm comb}^{{\rm BATSE}}=0.013$ $\pm 0.496$\\
&&  {\bf OSSE (64~ms) rescaled}\\
&& $\Delta t_1^{{\rm OSSE}}=-0.049$ $\pm 0.019$;
 $\Delta t_2^{{\rm OSSE}}=-0.046$ $\pm 0.019$;
 $\Delta t_3^{{\rm OSSE}}=-0.045$ $\pm 0.019$\\
&& {\bf combined BATSE+OSSE}\\
&& $\Delta t_{\rm comb}^{\rm BATSE+OSSE}=-0.047$ $\pm 0.011$\\
\hline
990308 (7457)&
1.2&  {\bf BATSE (64~ms)}\\
&& $\Delta t_1^{{\rm BATSE}}=0$ $\pm 0.860$;
$\Delta t_2^{{\rm BATSE}}=-0.064$ $\pm 0.860$;
$\Delta t_3^{{\rm BATSE}}=-0.256$ $\pm 0.860$\\
&& $\Delta t_4^{{\rm BATSE}}=-1.024$ $\pm 0.860$;
$\Delta t_5^{{\rm BATSE}}=0$  $\pm 0.860$;
$\Delta t_6^{{\rm BATSE}}=0.064$ $\pm 0.860$\\
&& $\Delta t_7^{{\rm BATSE}}=0$ $\pm 0.860$\\
&& {\bf combined BATSE (64~ms)}\\
&& $\Delta t_{\rm comb}^{{\rm BATSE}}=-0.183$ $\pm 0.325$\\
\hline
990510 (7560)&
1.619& {\bf BATSE (64~ms)}\\
&& $\Delta t_1^{{\rm BATSE}}=0.384$ $\pm 0.860$;
$\Delta t_2^{{\rm BATSE}}=0.448$ $\pm 0.860$;
$\Delta t_3^{{\rm BATSE}}=0$  $\pm 0.860$\\
&& $\Delta t_4^{{\rm BATSE}}=-0.256$  $\pm 0.860$;
$\Delta t_5^{{\rm BATSE}}=0$  $\pm 0.860$;
$\Delta t_6^{{\rm BATSE}}=-0.528$  $\pm 0.860$\\
&& $\Delta t_7^{{\rm BATSE}}=-0.128$  $\pm 0.860$;
$\Delta t_8^{{\rm BATSE}}=-0.256$  $\pm 0.860$;
$\Delta t_9^{{\rm BATSE}}=0$  $\pm 0.860$\\
&& $\Delta t_{10}^{{\rm BATSE}}=-0.448$  $\pm 0.860$\\
&& {\bf combined BATSE (64~ms)}\\
&& $\Delta t_{\rm comb}^{{\rm BATSE}}=-0.078$ $\pm 0.272$\\
&&  {\bf OSSE (64~ms) rescaled}\\
&& $\Delta t_1^{{\rm OSSE}}=-0.045$ $\pm 0.019$;
$\Delta t_2^{{\rm OSSE}}=-0.032$ $\pm 0.019$;
$\Delta t_3^{{\rm OSSE}}=-0.041$ $\pm 0.019$\\
&& {\bf combined BATSE (64~ms)}\\
&& $\Delta t_{\rm comb}^{{\rm BATSE+OSSE}}=-0.039$ $\pm 0.011$\\
\hline
991216 (7906)&
1.02& {\bf BATSE (64~ms)}\\
&& $\Delta t_1^{{\rm BATSE}}=-0.064$ $\pm 0.860$;
$\Delta t_2^{{\rm BATSE}}=-0.064$ $\pm 0.860$;
$\Delta t_3^{{\rm BATSE}}=-0.064$  $\pm 0.860$\\
&& $\Delta t_4^{{\rm BATSE}}=-0.064$  $\pm 0.860$;
$\Delta t_5^{{\rm BATSE}}=0.064$  $\pm 0.860$;
$\Delta t_6^{{\rm BATSE}}=0$  $\pm 0.860$\\
&& $\Delta t_7^{{\rm BATSE}}=0$  $\pm 0.860$;
$\Delta t_8^{{\rm BATSE}}=0$  $\pm 0.860$\\
&& {\bf combined BATSE (64~ms)}\\
&& $\Delta t_{\rm comb}^{{\rm BATSE}}=-0.040$ $\pm 0.304$\\
\hline
\end{tabular}\par \vspace{0.3cm}
\end{center}
\caption{ \it
Data on the light curves for GRBs with known redshifts used in this
analysis. The third column gives the time lags between the arrivals of
every identical genuine variation points in the third (high-energy) spectral
band and the first (low-energy) one. For every light curve, we give the weighted means of the time lags in $64$~ms BATSE domain, combining the genuine
variation points for that GRB. Both statistical and systematic errors
are included. The average spread of individual time lags is below the mesurement uncertainty. Also indicated are the results obtained by combining OSSE
$64$~ms light
curves in the $3$--$6$~MeV energy range with BATSE light curves in the
$115$--$320$~keV energy band. In the case of GRB980329, the results of BATSE
$64$~ms and TTE $2.7$~ms resolution measurements are combined with the
OSSE-BATSE $64$~ms comparisons.}
\label{table1}
\end{table*}

 \section{Analysis of Time Lags in Emissions from GRBs with Known
Redshifts}

Once the BeppoSAX satellite began to localize long bursts in the sky to
within a few arcminutes, and distribute their locations to observers
within hours, it turned to be possible to discover X-ray, optical, and
radio afterglows (Costa \etal \ 1997; van~Paradijs \etal \ 1997; Frail \etal \ 1997), and host galaxies. Subsequent observations
led to the spectroscopic determination of GRB redshifts, using absorption
lines in the spectra of the afterglows and emission lines in the spectra
of the host galaxies. By now, redshifts have been measured for about 20
bursts (see for example 
(Norris \etal \ 2000; {\tt http://www.aip.de/${\tilde {~}}$jcg/grbrsh.html}; Amati \etal \ 2002) and references therein).

Our first aim is to measure the timings of genuine variation points,
characterized by Lipschitz exponents as discussed above, for different
spectral bands in the light curves of distant GRBs. Correlating the
time lags between different energies with the GRB redshifts, we then try
to extract time delays related to the refractive index that may be induced
by quantum gravity.

We use genuine variation points with the same Lipschitz exponents
$\alpha$, measured in different energy bands, and assume that any initial
relative time lags attributable to the properties of source are
independent of redshift.
Thus, the key to disentangling quantum-gravity effects is reduced to the
problem of detecting genuine variation points with the highest possible
precision. The biggest uncertainties in our analysis come from our
procedure for estimating the DWT intensity profiles, whilst the errors
generated by the CWT zoom are negligible. The error in the
wavelet-shrinkage procedure is defined by the time-bin resolution in the
analysis of the light curve and the support of the DWT mother wavelet.

As shown in Table~\ref{table1}, we use GRB light curves from BATSE, which
have been recorded with 64~ms temporal resolution in four spectral
channels. Unfortunately, the BATSE catalog of light curves includes only
about a half of the GRBs with known redshifts. The light curves of the
other GRBs with known redshifts have been collected by other satellites
(BeppoSAX, HETE), and the data are not available publicly.  The BATSE
lower-level discriminator edges define the channel boundaries at
approximately 25, 55, 115 and 320~keV. We look for the spectral time lags
of the light curves recorded in the $115-320$~keV energy band relative to
those in the lowest $25-55$~keV energy band, providing a maximal lever arm
between photon energies. We do not use the fourth BATSE channel with
energies between $320$~keV and $1.9$~MeV, as these have ill-defined
energies and poorer statistics
- see Fig.~\ref{batse4}. Instead, we compare the
rather more energetic light curves accumulated by OSSE~\footnote{We are
grateful to M. Strikman for kindly providing us with OSSE data.} with the
$115-320$~keV BATSE light curves, which increases the lever arm for
probing photon propagation into the MeV range.

Since we apply the CWT zoom technique to detect the genuine variation
points of the reconstructed intensity profiles, we impose some conditions
on the choice of shrinking wavelet~\footnote{In most cases, discrete
wavelets cannot be represented by an analytical expression or by the
solution of some differential equation, and instead are given numerically
as solutions of functional equations (e.g. Mallat \ 1998).}. In general, when
choosing the appropriate wavelet basis, one has to strike a balance
between the degree of regularity of the wavelet, the number of its
vanishing moments $p$ and the size of its support. It is clear that the
size of the support defines uncertainties in the positions of genuine
variation points after the reconstruction of intensity profile. This
consideration motivates using the DWT basis with the most compact support
for the wavelet shrinkage procedure. On the other hand, to preserve
maximally the regularity of the original signal, one should use wavelets
with a high degree of regularity. In addition, one should avoid disturbing
significantly the alignment of peaks of the original light curves, which
motivates using symmetrical discrete wavelets.

The discrete wavelet that best reconciles the above requirements is that
called Symmlet-$p$ (e.g. Mallat \ 1998). It is the most symmetric, regular discrete
wavelet with minimum support. The number $p$ of vanishing moments defines
the size of the support, and consequently the errors of the position
estimations $\frac{1}{2}(2p-1)\times {\rm bin-size}$. Moreover the same
number $p$ of vanishing moments defines the regularity of Symmlet-$p$. For
a large number of vanishing moments, the Lipschitz regularity of
Symmlet-$p$ is $0.275p$ (Daubechies \ 1991). Thus, to have more then $2$ continuous
derivatives, $p$ should exceed $8$.

For the selected GRBs in Table~\ref{table1}, we have performed the wavelet
shrinkage procedure using a Symmlet-$10$ basis~\footnote{The coefficients
of the Symmlet filters are tabulated in WAVELAB toolbox \\ ({\tt 
http://www-stat.stanford.edu/${\tilde {~}}$wavelab}), for example.}.
At sufficiently high signal-to-noise ratio levels, this procedure tends to
preserve the regularity of the light curves. In some cases, namely for
GRB980329 and GRB970508, we applied the translation-invariant (Mallat \ 1998)
version of the shrinkage procedure with reduced threshold. This procedure
implies averaging estimates produced from the original signal itself and
from all shifted versions of the signal, and allows one to
avoid artefacts while preserving the real transient structure.
Subsequently, we apply the CWT zoom technique for reconstructing intensity
profiles to identify the arrival times of genuine variation points and
estimate their Lipschitz exponents in every spectral band. We consider
that a genuine variation point has been detected if it has Lipschitz
exponent $\alpha\le 1$. Genuine variation points found in the vicinity of
each other, but belonging to two different spectral bands, are considered
to have been generated at the source if the values of their Lipschitz
exponents are equal to each other. The other variation points with
$\alpha$ substantially exceeding $1$ exhibit only smooth transitions of
the signal, and do not mark sharp transient time structures. We recall
that only sharp transient structures are important in the search for
spectral time lags.
For seven GRBs out of nine, we detected more than one pair of identical genuine variation
point per light curve, as seen in Table~\ref{table1}. The systematic
errors (Kolaczyk \ 1997) were estimated by using different resolution levels
($L=6,\ 5,\ 4$) in the wavelet shrinkage procedure.

In order to probe the energy dependence of the velocity of light that
might be induced by quantum gravity, we have compiled the whole available
data in Table~\ref{table1} as functions of the variables $K_{\rm l}$ and $K_{\rm q}$,
defined by the integrals in \form{timedel1} and \form{timedel2},
respectively. In the case of linear quantum-gravity corrections, the
variable takes the form

\beq
\label{K1}
K_{\rm l}=\int\limits_0^z\frac{dz}{h(z)},
\eeq
whilst for the quadratic case we use
\beq
\label{K2}
K_{\rm q}=\int\limits_0^z\frac{(1+z)dz}{h(z)}.
\eeq
Since both \form{timedel1} and \form{timedel2} exhibit linear dependences
on the variables \form{K1} and \form{K2} respectively, we perform a
regression analysis for a linear dependence of the time lags between pairs
of genuine variation points, in the form
\beq
\label{regreq}
\Delta t=aK+b.
\eeq
The result of our
regression fit to the full 64~ms statistics for linear quantum-gravity
corrections \form{linearqg} is shown in Fig.~\ref{regr64}.
The best fit corresponding to Fig.~\ref{regr64} is given by
\beq
\label{fir64l}
\Delta t=0.60(\pm 0.46)K_{\rm l}-0.72(\pm 0.53).
\eeq
Following the same procedure for the quadratic corrections, one gets
\beq
\label{fir64q}
\Delta t=0.17(\pm 0.17)K_{\rm q}-0.42(\pm 0.32).
\eeq
More precise results can be obtained by combining BATSE and OSSE data.
Four light curves accumulated by OSSE exibit structures that
can be compared with similar features observed by BATSE, as seen in
Table~\ref{table1}. Since the OSSE data are at higher energies:
$0.15-10$~MeV,
one has to rescale the results of OSSE-BATSE comparison in order to
combine them directly with results obtained using the third BATSE channel,
by a factor:
\beq
\label{rescale}
\frac{110~{\rm keV}-55~{\rm keV}}{3~{\rm MeV}-110~{\rm keV}},
\eeq
where $3~{\rm MeV}$ is the energy at which the contribution
of flux accumulated by OSSE becomes significant. The spectral information
for GRB980123 indicate that half of the total flux has been accumulated in
the energy range $3-6$~MeV.

One may also increase the sensitivity of the determination of time
lags by using BATSE time-tagged event (TTE) data, which record the arrival
time of each photon with a precision of $2~\mu$s, in the same four
energy channels. The onboard memory was able to record up to 32,768
photons around the time of the BATSE trigger. Typically, this quota of
photons was filled in 1 or 2~s. For short GRBs, the mean structure of the
whole light curve might be in the TTE data, along with substantial periods
of background emission after the burst, whilst for the long-duration GRBs
that we analyze, as in Table~\ref{table1}, the TTE data cover only the
leading portion of light curve. We have rebinned with resolution $\simeq
1$~ms the leading portions of all the GRBs from Table~\ref{table1}
using TTE data. Only one light curve, that of GRB980329, yields a signal
with clearly identified isolated singularities in the first and third
spectral bands.  The statistics available to detect genuine variation
points in this light curve yield a resolution of 2.7~ms. Combining
this TTE measurement with the 64~ms BATSE measurements and the results of
our BATSE-OSSE comparisons, we get the following results:
\beq
\label{lintte}
\Delta t=0.010(\pm 0.022)K_{\rm l}-0.053(\pm 0.026)
\eeq
\beq
\label{quadrtte}
\Delta t=0.003(\pm 0.006)K_{\rm q}-0.048(\pm 0.016)
\eeq
for linear Fig.~\ref{regrtte} and quadratic corrections respectively.
The single BATSE point with much higher precision than the others does not
improve substantially the significance of the fits. However, including the
OSSE-BATSE
and TTE measurements into the overall fit does improve the sensitivity
(see the next section) and makes the result less dependent on the
properties of individual sources.

The leading parts of other light curves from Table \ref{table1}. which do
not exhibit coherently variable structures, can be characterized as
fractal signals without isolated singularities. One can also analyze such
singularities with CWT (e.g. Dremin \etal \ 2001; Mallat \ 1998), but such a study lies beyond
the scope of this paper.

\section{Compilation of Limits on Quantum Gravity}
We now analyze the likelihood function to derive the results of our search
for a vacuum refractive index induced by quantum gravity. We establish a
95~\% confidence-level lower limit on the scale $M$ of quantum gravity by
solving the equation

\beq
\label{likl}
\frac{\int\limits_{M}^{\infty}L(\xi )d\xi}{\int\limits_{0}^{\infty}L(\xi )d\xi}=0.95
\eeq

\noindent where $\infty$  symbolizes a reference
point fixing the normalization. In our case, we choose as reference point
${\cal M} = 10^{19}$~GeV, the Planck mass, which corresponds to the
highest possible scale above which quantum-gravity effects vanish and
corrections to the vacuum refractive index become infinitesimally small. 
In practice variations of reference point, even by an order of magnitude, does not influence the final result. 

We use the fact that only the coefficient $a$ in \form{regreq} is related
to the quantum-gravity scale, whereas $b$ includes a possible unknown
spectral time lag inherited from the sources, which we assume to be
universal for our data set. With this assumption, one can shift our data
points by an amount $-b$, taking $b$ from the best fits \form{fir64l} and
\form{fir64q}, and use $L\propto\exp (-\chi^2(M)/2)$ (with normalisation
appropriately fixed to unity) as the likelihood function in \form{likl},
where \beq \label{chi2} \chi^2(M)=\sum\limits_{\cal D}\left[\frac{\Delta
t_i-b_{\rm shift}-a(M)K_i}{\sigma_i}\right]^2 \eeq is calculated for all the
possible values of $a(M)$, defined by the coefficients of $K_{\rm l}$ ($K_{\rm q}$) in
\form{timedel1} and \form{timedel2}, respectively. The sum in \form{chi2}
is taken over the all the data points $\Delta t_i$, which are symbolized
by ${\cal D}$, with $\sigma_i$ characterizing the uncertainties in the
measured time lags. The calculated $\chi^2(M_{\rm L})$ for the different
combinations of the data sets we use in the case of linear corrections to
the refractive index \form{timedel2} is shown in Fig. \ref{likelihood}.
The minima of $\chi^2$ correspond the `signal-like' regions, where the
data are better described by a scenario with a refractive index, induced
by quantum gravity. The most robust estimation on the lower limit of
quantum gravity with a linearly energy-dependent correction is obtained
from the combination of all the data sets, and is indicated by a solid
line in Fig.~\ref{likelihood}: 
\beq 
\label{linlimit} 
M_{\rm L}\ge 6.9 \cdot 10^{15}~{\rm GeV} 
\eeq 
Similar considerations lead to the 
following lower
limit on quadratic quantum-gravity corrections \form{quadraticqg} to the
photon dispersion relation:
\beq
\label{quadlimit}
M_{\rm Q}\ge 2.9\cdot 10^6~{\rm GeV}.
\eeq
To the accuracy stated, we find identical numerical results, whether we 
use a logarithmic 
measure for $M$ cut off at ${\cal M} = 10^{18,19,20}$~GeV, 
as shown in Fig.~\ref{likelihood}, 
or a $1/M$ measure integrated to infinity, 
\beq
\label{invmeasure}
\frac{\int\limits_{0}^{1/ M}L'(\xi 
)d\xi}{\int\limits_{0}^{\infty}L'(\xi )d\xi}=0.95,
\eeq
where $L'(1/M)$ is the likelihood function with respective to $1/M$, as
shown in Fig.~\ref{likelihoodinv}.

The key result (\ref{linlimit}) is significantly stronger than that 
in (Ellis \etal \ 2000b), thanks to the improved analysis technique using wavelets 
and the use of a more complete dataset.

\section{Discussion}

We have investigated in this paper possible non-trivial properties of the
vacuum induced by quantum gravity, by probing modifications of the
dispersion relation for photons. These features can appear in several
approaches to quantum gravity, including Liouville string theory, models
with large extra dimesions (Campbell-Smith \etal \ 1999) and loop gravity. Similar
modifications of dispersion relations can take place for the other matter
particles, leading to other non-trivial effects such as changes in the
thresholds for some reaction attenuating ultra-high energy cosmic rays
(UHECR), or vacuum \v{C}erenkov-like radiation (see (Sarkar \ 2002) and
references therein), which could have a large influence on the
interpretation of the puzzling astrophysical data on UHECR. 

We have attempted to extract the most complete information about the
possible vacuum refractive index induced by quantum gravity, by using
wavelets to look for any correlation with redshift of the time lags
between the arrival times of sharp transients in GRB light curves observed
in high- and low-energy spectral bands. This analysis combined continuous
wavelet transforms to remove noise and discrete wavelet transforms to
identify sharp transients in different spectral bands. Eight GRBs with
known cosmological redshifts and light curves available publicly have been
used in our analysis.

 It is instructive to compare the time lags we find with those found using cross-correlation
analysis (Band \ 1997;  Norris \etal \ 2000; Norris \ 2002). 
Our measurements of the spectral time lags in BATSE $64$~ms
light curves for five GRBs, which are common for the sample we used and that under consideration in Norris \etal \ 2000, are 
consistent in absolute value with the trend found in Norris \etal \ (2000) for bursts with
higher luminosities (which are closer, on the average) to have shorter
time lags. 
Discrepancies are found in two cases out of five common GRBs. Namely, we found soft-to-hard evolution for GRB971214 and GRB990123 (positive time lag), which is opposite to the hard-to-soft evolution (negative time lag) found in Norris \etal \ 2000. These two GRBs havea complicated structure of emission: GRB971214 has a wide ``clump'' of emission which consists of spikes that are barely overlapped, while GRB990123 consists; in two intensive wide pulses with a quite complicated cluster afterwards. Thus these two GRBs can be attributed to a lack of morphological classification power of the cross-correlation technique due to the problem of interpretation of the CCF pike width (Band \ 1997). As an explicit example, we analysed GRB941119, which has been assigned, in Band \ 1987, as the GRB without clear spectral evolution with the respective to cross-correlation analyses. The light curve of GRB941119 consists in cluster with several closely spaced spikes protruding from a smooth envelope. We found five pairs of identical genuine variation points in first and third spectral bands, demonstrating all together the soft-to-hard spectral evolution with weghted average time lag $\Delta t=0.051\pm 0.384$~s.  One can see from Table \ref{table1} that in several cases different pairs of identical genuine variation points detected in  one and the  same GRB's light curve demonstrate different kinds of spectral evolution, namely either hard-to-soft or soft-to-hard. This fact could be connected with a possible intrinsic spectral evolution during the burst progress. The example of GRB941119 is one of such cases, demonstrating unclear spectral evolution with the respect to the cross-correlation analysis. The  wavelet technique we apply classifies more explicitly a variable spectral evolution and consequently gives more accurate results for the weighted average spectral evolution, as in case of GRB941119. 
The time lags measured
between the BATSE 25-55~keV and OSSE 3-6~MeV light curves exceed
substantially the time lags between the first and third BATSE energy
bands. Without the correction by the ratio \form{rescale}, the BATSE-OSSE
time lags we find are $\simeq -2$~s, in good agreement with Piro \etal 1998,
where the time lag between bands of GRB970228 with a similar energy
difference was estimated using BeppoSAX data. The wavelet technique is very effective also to deal with transient signals such as TTE data with low signal-to-noise ratios, where cross-correlation analyses can fail. 

We believe that  there is no regular cosmological evolution of the
sources we use. This fact is already widely accepted in the literature on other high-redshift
sources, namely supernovae. So there is no effect that can cause any correlation of weighted average time lags with redshifts of the sources. Thus any correlation of spectral time lags we may find can be attributed to the effects of propagation.

Finally it should be stressed that the method we applied is very powerful in the sense of analysing signals with strong non-linear dynamics behind, as GRBs could be. Of course it does not pretend to explain the underlying dynamics and physical origin of GRBs, but it gives a hint  of the stage at which the most variable processes take place and characterize quantitatively the degree of instability accompanying those processes. In our case the radiation from genuine variation points is considered as the messenger of a fast non-linear dynamics at the source. The Lipschitz exponents, which characterize  quantitatively the `degree' of instability of the dynamics, give the information of whether photons of different energies have been produced at one and the same event at the source. So this gives an ideal 'time offset' between energy bands to measure any differences in the speed of light.  

It is widely accepted (see Band \etal \ 1997; Norris \etal \ 2000, and references therein)  
that the spectral evolution in GRBs leads to peaks migrating later in
time. These time lags are not directly connected with the distance to the
source, but are correlated with intrinsic properties of GRBs, such as
luminosity (Norris \etal \ 2000) or variability (Shaefer \etal \ 2001). The quantum-gravity
energy-dependent time delay plays the role of a foreground effect of
opposite sign to the usual spectral evolution of GRBs, which increases
with distance. Hence model-independent information about quantum gravity
can be extracted only from a statistical analysis of sources with a known
distance distribution.

We have not found a significant correlation of the measured time lags with
the cosmological redshift that would indicate any deviation of the vacuum
refractive index from unity. This fact allows us to establish 95 \% C.L.
lower limits on the quantum-gravity scale at the levels $6.9\cdot
10^{15}$~GeV and $2.8\cdot 10^{6}$~GeV for linear and quadratic 
distortions
of the dispersion relation, respectively. However, despite the lack of
any significant evidence for a quantum-gravity signal, there is a region
of the linear scale parameter $M$ where the data are better described by a
scenario with a refractive index induced by quantum gravity. This fact
indicates that any increase in the statistics, especially with higher
resolution, would be of the utmost interest for exploring the possibility
of a quantum-gravity correlation of spectral time lags with redshift. For
this reason, we urge the light curves for all GRBs with measured redshifts
to be made generally available, as is already the case for BATSE data.

It has been observed that, if the dispersion laws for elementary
particles differ from the standard ones, the expansion of the Universe may
result in the gravitational creation of pairs of particles and
antiparticles with very high energies (Starobinsky \& Tkachev \ 2002). The expansion of the
Universe (both at present and in the early Universe) gradually redshifts
Fourier modes of a quantum field, and may transport them from the
trans-Planckian region of very high momenta to the sub-Planckian region
where the standard particle interpretation is valid. Then, if the WKB
condition is violated somewhere in the trans-Planckian region, the field
modes enter the sub-Planckian region in a non-vacuum state containing
equal numbers of particles and antiparticles. The most restrictive upper
limit follows from the number of UHECR created at the present
epoch (Starobinsky \& Tkachev \ 2002). This limit, together with our measurements of the vacuum
refractive index, may rule out the possibility of detecting imprints
of trans-Plankian physics on the CMB anisotropy, as proposed
in (Brandenberger \& Martin \ 2002).

It is a widespread belief that the combination of quantum theory with
gravity is an explosive area, to the extent that it blows up the concrete
fabric of space-time, splitting it into space-time foam. It is true that,
as yet, there is no universally accepted mathematical model of this
metamorphosis, while several attempts at it are characterized by a varying
level of success. However, despite the firm and widely-held opinion that
quantum gravity defies experimental tests, it is remarkable that several
such tests have been proposed in recent years. Some of them have put
rather severe constraints on some quantum-gravity models, whilst some
models based on non-critical strings seem to remain unscathed so 
far (Ellis \etal \ 2002; Ellis \etal \ 2002a). In
particular, as we have shown in this paper, using the most sophisticated
technique available, that of wavelets, to pick up authentic time-lag
effects in gamma-ray bursts, we have been able to put a rather firm lower
bound on the dispersion of light in the vacuum: $M > 6.9 \times
10^{15}$~GeV. This approaches the range of scales where we might expect
such an effect to appear. Data able to test further this possibility
already exist, and more data will soon be available. 

{\it Une affaire \`a suivre ...}.

\begin{acknowledgements}

This work is supported in part by the European Union through contract
HPRN-CT-2000-00152. The work of D.V.N. is supported by D.O.E. grant
DE-FG03-95-ER-40917. We are grateful to A.~Biland for his kind help for
reading out the TTE data. We thank Hans Hofer for his continuing support and
interest. We are also greateful to I.M.~Dremin for useful conversations on
wavelets, and A.S.S. is indebted to E.K.G.~Sarkisyan for many useful
conversations on statistics.

\end{acknowledgements}

\appendix \section{Wavelet-Based Thresholding Estimator}

A thresholding estimator of a $n=2^J$-component discrete signal $X[n]$
decomposed at the resolution level $L$ (see Section 4) in a wavelet basis
can be written (Mallat \ 1998) as:

\beq
\label{treshestimator}
\tilde F=\sum\limits_{j=L}^{J-1}\sum\limits_{m=0}^{2^{j-1}}\rho_T(d_{j,m}^X)
\psi_{j,m}+\sum\limits_{m=0}^{2^{L-1}}\rho_T(c_{J,m}^X)\phi_{J,m},
\eeq
where the function $\rho_T$ provides a soft threshold:
\beq
\label{softthreshold}
\rho_T(x)=\left\{\begin{array}{lcl}
x-T&{\rm if}&x\ge T\\
x+T&{\rm if}&x\le -T\\
0&{\rm if}&|x|\le T
\end{array}
\right.
\eeq
If the signal $X[n]$ is contaminated by additional noise $W[n]$,
the threshold $T$ is generally chosen so that there is a high probability
that it is just above the maximal level of this noise. Since $W[n]$ is a
vector of $N$ independent Gaussian random variables of variance
$\sigma^2$, one can prove (Mallat \ 1998) that the maximum amplitude of the
noise has a very high probability of being just below $T=\sigma\sqrt{2\log
n}$.

A signal $X[n]$ of size $n$ has $n/2$ wavelet coefficients $\{
d_{J-1,m}\}_{0\le m< n/2}$ at the finest scale. The coefficients of $f$
itself, $|d_{j,m}^f|$, in the sum \form{contamination} are small if $f$ is
smooth over the support of $\psi_{J-1,m}$, implying $d^X_{J-1,m}\approx
d^W_{J-1,m}$. In contrast, $|d^f_{J-1,m}|$ is large if $f$ has a sharp
transition in the support of $\psi_{J-1,m}$.  A piecewise-regular signal
has few sharp transitions, and hence produces a number of large
coefficients that is small compared to $n/2$. At the finest scale, the
signal $f$ thus influences the value of a small portion of large-amplitude
coefficients $d_{J-1,m}$, that are considered to be `outliers'. All the
others are approximately equal to $d_{J-1,m}^W$, which are independent
Gaussian random variables of variance $\sigma^2$.

A robust estimator of $\sigma^2$ is calculated from the median of
$(d_{J-1,m}^X)_{0\le m< n/2}$. The median of $P$ coefficients
${\rm Med}(\alpha_p)_{0\le p< P}$ is the value of the middle coefficient
$\alpha_{n_0}$ of rank $P/2$. As opposed to an average, it does not depend
on the specific value of coefficients $\alpha_p >\alpha_{n_0}$. If $M$ is
the median of the absolute value of $P$ independent Gaussian random
variables of zero mean and variance $\sigma_0^2$, then one can
show (Mallat \ 1998) that the expected value of $M$ is $0.6745\sigma_0$. Thus
the variance $\sigma^2$ of the noise $W$ is found from the median $M_X$ of
$(|d_{J-1,m}^X|)_{0\le m< n/2}$ by neglecting the influence of $f$:
\beq
\label{noisenorm}
\tilde\sigma =\frac{M_X}{0.6745}.
\eeq
One may say that $f$ is responsible for few large amplitude outliers,
and that these have little impact on $M_X$. In practice, it is more
convenient to transform the signal, by scaling it in such a way that the
wavelet coefficients at the finest level of decomposition have median
absolute deviation equal to unity, as seen in Fig~\ref{smoothed}. 
\label{apa}.

\section{Lipschitz Regularity}

Taylor expansion relates the differentiability of a signal to local
polynomial approximations. Suppose that $f$ is $m$ times differentiable in
$[\nu -h;\nu +h]$. Let $p_{\nu}$ be the Taylor polynomial in the
neighborhood of $\nu$. The $m^{th}$ order of differentiability of $f$ in
the neighborhood of $\nu$ yields an upper bound on the error
$\epsilon_{\nu}(t)=f(t)-p_{\nu}(t)$ when $t$ tends to $\nu$. The Lipschitz
regularity extends this upper bound to non-integer exponents. Namely, a
function $f$ is pointwise Lipschitz: $\alpha\ge 0$ at $\nu$, if there
exists a polynomial $p_{\nu}$ of degree at most $\alpha$ such that

\beq
\label{lip}
|f(t)-p_{\nu}(t)|\le {\cal K}|t-\nu |^{\alpha}
\eeq
where ${\cal K}$ is a constant.  The polynomial $p_{\nu}(t)$ is uniquely
defined at each $\nu$. If $f$ is $m\le\alpha$ times continuously
differentiable in a negborhood of $\nu$, then $p_{\nu}$ is the Taylor
expansion of $f$ at $\nu$. Pointwise Lipschitz exponents can vary
arbitrarily from abscissa to abscissa. If $f$ is uniform Lipschitz:
$\alpha >m$ in the neighborhood of $\nu$, then one can verify that $f$ is
necessarily $m$ times continuosly differentiable in this neighborhood. If
$\nu\le\alpha <1$, then $p_{\nu}(t)=f(\nu )$, and the Lipschitz condition
\form{lip} becomes

\beq
\label{lip1}
|f(t)-f(\nu )|\le {\cal K}|t-\nu|^{\alpha},
\eeq
$f$ is not differentiable at $\nu$ and $\alpha$ characterizes the
singularity type. For more references to the mathematical 
literature, see (Dremin \etal \ 2001; Mallat \ 1998).
\label{apb}

\end{document}